# The Effect of Tissue Detection on False Positives of Diffusion-Based Artifact Detection in Histopathology

Konstantinos Moutselos*, Ilias Maglogiannis

*Department of Digital Systems, University of Piraeus, Piraeus, Greece*

*Corresponding author: Konstantinos Moutselos, Department of Digital Systems, University of Piraeus, 80 Karaoli & Dimitriou St., 18534 Piraeus, Greece. E-mail: kmouts@unipi.gr.

ORCID: K. Moutselos 0000-0002-6759-8540; I. Maglogiannis 0000-0003-2860-399X.

## Declaration of competing interest

The authors declare that they have no known competing financial interests or personal relationships that could have appeared to influence the work reported in this paper.

## Funding

This work was co-financed by the European Union and national funds through the National Strategic Reference Framework (NSRF) 2021–2027, under the Operational Programme "Competitiveness", Action "Bilateral Scientific and Technological Cooperation Greece–China" (Call Code 12KE, State Aid Code 11434), within the project "Research of Digital Pathology Image-based Intelligence Diagnosis Techniques of Prostate Cancer and Development of DT Software System - PRODIGY" (Grant Code: ΔΣΕΚΡ01-0021403, ERDF / European Regional Development Fund).

## Abstract

One-class artifact detectors for whole-slide images learn normal tissue from a clean training pool and flag departures from it. The pool is built by a preprocessing pipeline whose tissue-detection step is usually treated as neutral. We tested whether it is. On 16 annotated slides from The Cancer Genome Atlas, we rebuilt the clean pool of a diffusion-based detector with different tissue detection methods and compared the resulting models in a four-fold cross-validation. Per-slide saturation–Otsu detection excluded normal tissue, chiefly tissue with large clear spaces such as adipose tissue and alveolar parenchyma, and on slides with thick marker ink kept the ink while excluding ordinary tissue. Replacing it with entropy-based detection reduced the false-positive fraction on held-out clean slides from 0.102 to 0.016, in every fold and with a second training seed, without loss of sensitivity; the gain came from the composition of the pool, not its size. Across three tissue detection methods, false positives followed the fraction of such clear-space tissue in the pool, a statistic that needs no labels or training (0.103, 0.016 and 0.009). The effect did not carry over at the same size to a nearest-neighbour detector on foundation-model features. On an external cohort, the curated pool lowered clean-control false positives by about 20%, far less than on the development slides, and the remaining cross-center loss was not explained by stain differences. For one-class quality control, tissue detection decides what the model learns as normal and should be chosen and reported accordingly.

## 1. Introduction

Whole-slide images carry artifacts introduced during tissue processing and scanning: tissue folds, out-of-focus regions, air bubbles and marker ink, among others. Left undetected, they degrade downstream computational pathology models [1,2], and automated quality control has become a routine preprocessing step [3,4]. Regions it flags are either discarded, at the risk of losing diagnostic tissue, or restored by generative models [5]; either way, a false positive costs normal tissue. Supervised detectors require pixel-level annotations for every artifact class and generalize poorly to classes they were not trained on. One-class detectors avoid both limitations: they learn the distribution of normal tissue from a clean training pool and flag whatever departs from it. Reconstruction-based diffusion models are a recent instance of this approach in histopathology [6–8].

For a one-class detector, the clean training pool is not an incidental input: it is the definition of normal. It is usually assembled by a preprocessing pipeline that tiles the slide, keeps the tiles judged to be tissue, and removes those overlapping annotated artifacts. The tissue-detection step in that pipeline is commonly treated as neutral, a matter of discarding glass. Its effect has been studied mainly in two ways. Tissue detection methods are compared by their overlap with reference tissue masks [9], and the few studies of downstream effects concern supervised models: in prostate cancer grading, replacing thresholding-based with deep-learning tissue detection did not significantly change overall Gleason grading performance, although it reduced complete detection failures [10]. For a supervised model, tissue detection decides which regions are scored. For a one-class model, it also decides what is learned as normal: tissue excluded from the pool lies outside the support of the learned distribution and is flagged as anomalous when it appears at test time [11]. How much of a one-class detector's false-positive burden originates in the tissue detection used to build its pool has, to our knowledge, not been measured.

We encountered the question in our own work on a diffusion-based artifact detector. After reproducing and extending a published detector [12,13], a decomposition of its benchmark performance [14] showed that its false positives on clean tissue concentrate in normal tissue with large clear spaces, such as adipose tissue and alveolar parenchyma, which the saturation-based tissue detection used to build the training pool classifies as background. We call this clear-space tissue: normal tissue in which much of the area is lumen, alveolar space or fat vacuole, bounded by thin walls or septa. The present study tests that observation as a causal claim.

We rebuilt the clean pool with different tissue detection methods and compared the resulting detectors in a four-fold cross-validation over 16 annotated slides from The Cancer Genome Atlas (TCGA). We then asked whether the effect is graded, whether it depends on the detector, and whether it transfers to an external cohort. Our contributions are:

1. We show that per-slide saturation–Otsu tissue detection systematically excludes normal clear-space tissue from the clean pool and, on slides with thick marker ink, keeps the ink while excluding ordinary tissue.
2. We show that replacing it with entropy-based tissue detection removes most false positives on held-out clean slides (a more than sixfold reduction, reproduced with a second training seed and in every fold) without loss of sensitivity, and that the gain comes from the composition of the pool rather than its size.
3. We show that the effect is graded: across three tissue detection methods, clean-slide false positives follow the fraction of clear-space tissue each method admits to the pool, a statistic that needs no labels and no training.
4. We delimit the claim: the effect does not carry over at the same size to a nearest-neighbour detector on foundation-model features; on an external cohort it is in the expected direction but not confirmed, and the remaining cross-center loss is not explained by stain differences.
5. We release the pools, fold definitions, code and evaluation outputs, and give practical recommendations for building clean pools for one-class quality control.

The remainder of the paper is organized as follows. Section 2 describes the data, detector, pools and evaluation design; Section 3 reports the results; Section 4 discusses their implications and limitations; Section 5 concludes.

## 2. Materials and methods

### *2.1. Data*

**Development set.** We used the 16 training slides of the AIRAQC benchmark [15], all from TCGA: nine lung (adenocarcinoma and squamous cell carcinoma), three breast, two stomach, one cervix and one frozen normal-tissue section (TCGA-W5-AA30, sample type 11A; the only slide not formalin-fixed and paraffin-embedded). The AIRAQC annotations mark out-of-focus regions, tissue folds, penmarks and air bubbles within the tissue; by design they do not mark artifacts on the surrounding glass [15]. Four of the slides carry no annotation above 0.01% of their pixels and serve as clean slides (Section 2.4). Because the backbone of the detector was pretrained on a corpus that includes TCGA [16], all results on these slides are development-tier evidence; external evidence comes from SlideInspect [17]. No slides were scanned for this study: the TCGA slides are distributed as Aperio SVS files, and SlideInspect was assembled from several centers digitized with Aperio, Hamamatsu and 3DHistech scanners.

**External set.** We used the public test release of SlideInspect (Kaggle, artemis90/slideinspect), which provides 441 images, each with a manual artifact mask (out-of-focus regions, mounting artifacts and folds) and a manual tissue mask. The dataset authors confirmed that the released images contain no material from TCGA or The Cancer Imaging Archive, which places them, to our knowledge, outside the backbone's pretraining corpus, that only the 5× versions exist, and the encoding of the three artifact values (out-of-focus, mounting artifact, fold). Keeping H&E stains only (152 immunohistochemistry, periodic acid–Schiff and trichrome images excluded) leaves 289 images, of which 55 carry no artifact annotation (clean controls) and 135 contain out-of-focus regions; in six further images the annotation vanishes at the scoring resolution, and these are neither controls nor counted for sensitivity. "Mounting artifact" is a broader category than air bubble and is reported under its own name. Artifact pixel mass is concentrated on few images, so all comparisons are paired and made at the image level. The dataset is used under its academic, non-commercial terms. The images are stored at 5×; for inference, each 512-pixel crop is upsampled twofold (bilinear) to the model's 1,024-pixel window at the 10×-equivalent scale, and tissue is detected by the same saturation-based thumbnail mask as on the development slides. The manual tissue mask is not used for ingestion because it excludes out-of-focus tissue by definition.

### *2.2. Detector*

The detector is a one-class diffusion model: it learns to denoise patches of normal tissue and flags regions it denoises poorly. We used the PixCell diffusion transformer [16] with low-rank adapters (rank 16, alpha 16) [18] on its transformer blocks, trained with the conditioning-noise recipe of [13] 1,000 steps, batch size 4, learning rate $1\times10^{-5}$, training seed 42, conditioning drawn at each step from the statistics of UNI2-h embeddings [19] of the original pool, and no contrastive term. Training uses clean patches only; artifact annotations never enter the loss.

Patches are 1,024 × 1,024 pixels read at the pyramid level closest to 10× magnification, encoded by the frozen PixCell autoencoder into a 16-channel latent of 128 × 128 positions.

At inference, each slide is tiled into non-overlapping 1,024-pixel windows, keeping those in which at least 10% of the area is tissue on the saturation thumbnail mask; noise at diffusion step t = 800 is added to the latent, with the same noise seed for every model compared (42 on the development slides, 0 on SlideInspect), so that comparisons are paired in the noise draw; the per-position error is the squared difference between the predicted and the added noise, averaged over channels. Window errors are placed on a grid of one cell per 8 × 8 pixels and averaged where windows overlap. The reference model is the detector trained on the original pool [13].

**Thresholding rule.** Error maps are binarized without labels, per model. Two percentiles of the raw error, the 65th and 75th, are computed over the scored tissue of all slides being evaluated together (a fold's four held-out slides; all SlideInspect images). Each map is smoothed (Gaussian, σ = 2 cells), clipped to the range between the two percentiles, thresholded by Otsu's method within the covered area, and cleaned by a morphological closing followed by an opening with a 5 × 5 structuring element. The rule has no parameter tuned on the evaluation slides and is used throughout.

### *2.3. Clean pools and tissue detection*

**Candidate patches.** On each training slide we enumerate a regular grid of 1,024-pixel patches with a stride of 512 pixels and discard every patch whose box overlaps any artifact annotation. A patch enters the pool if at least 50% of its box is tissue according to the tissue detector, evaluated on the slide thumbnail (or, for GrandQC, on its mask) mapped to the patch coordinates.

**Tissue detection methods.** (i) *Saturation-based*: the saturation channel of the thumbnail in hue–saturation–value (HSV) space, thresholded per slide by Otsu's method. (ii) *Entropy-based*: the entropy method of LazySlide [20] (version 0.12.0). (iii) *GrandQC*: the public GrandQC masks for TCGA [4], taking classes 1–6 (normal tissue or tissue with an artifact) as tissue and excluding background and unclassified pixels. GrandQC provides no mask for the frozen slide TCGA-W5-AA30, whose 21 patches were kept as in the curated pool.

**Marker ink.** Marker ink was removed by visual review of contact sheets under an ink-only rule: patches with marker ink were excluded; other unannotated findings (dust, surgical-margin dye) were recorded but not removed, so that the pools differ only in tissue detection.

For the curated pool, all patches accepted by entropy-based detection were reviewed; for the GrandQC pool, patches shared with the curated pool inherit its review and the 373 patches new to GrandQC were reviewed separately (203 excluded).

**Pools.** The original pool (saturation-based, 2,185 patches), the curated pool (entropy-based, ink removed; 2,969) and the GrandQC pool (2,991).

**Clear-space fraction.** A label-free measure of how much clear-space tissue a pool contains: the mean over its patches of the fraction of pixels whose optical density, summed over the RGB channels, is below 0.15, computed on the patch downsampled to 256 × 256 pixels by box averaging. It needs no training.

### *2.4. Cross-validation design*

The 16 training slides were split into four folds of four held-out slides, fixed before any training: each fold holds out one slide carrying marker ink, one annotation-free clean slide and two others, and the two slides contributing most patches to the pools (TCGA-33-4532 and TCGA-55-A492) are in different folds. Air bubbles occur only in two folds, so the air-bubble sensitivity tolerance (Section 2.5) is checked on those two folds only. For each fold, a model is trained on the pool built from the remaining 12 slides and scored on the four held-out slides only. Before the first cross-validation run we verified that training from an explicit pool file reproduces the reference model's logged training losses exactly when given the original pool, so that differences between conditions arise from the data, not the code. The conditions were: original pool; original pool without marker ink (the final ink-exclusion list applied to the original pool, removing 157 patches); curated pool; curated pool with per-slide balancing (each patch's loss weighted by $n^{(a-1)}$, where n is its slide's patch count, with $a = 0.5$ and weights renormalized to mean 1, so that a slide's total weight grows as $\sqrt{n}$); curated pool randomly subsampled, per fold, to the size of the original pool without marker ink (seed 0, patch order kept); the original and curated pools with a second training seed (7); and the GrandQC pool. The primary comparison used the curated pool with per-slide balancing; once balancing was shown to add nothing, all later comparisons used the curated pool without it. Pool sizes per fold are given in Table 1.

**Table 1. Cross-validation folds and clean training pools.** Each fold holds out four slides (clean slide in bold) and trains on the pool built from the remaining 12. Pool sizes in patches of 1,024 × 1,024 pixels; clear-space fraction: mean fraction of near-white pixels per patch.

| **Fold** | **Held-out slides** | **Original** | **Original, ink removed** | **Curated** | **Curated, subsampled** | **GrandQC** | **Clear-space, original, ink removed** | **Clear-space, curated** | **Clear-space, GrandQC** |
|---|---|---|---|---|---|---|---|---|---|
| F1 | 49-4514, **55-A492**, 85-8052, HU-8610 | 1,701 | 1,594 | 2,004 | 1,594 | 2,083 | 0.095 | 0.146 | 0.181 |
| F2 | 21-1080, **55-8204**, 62-A46U, 86-8076 | 1,775 | 1,679 | 2,520 | 1,679 | 2,557 | 0.112 | 0.221 | 0.240 |
| F3 | **A8-A093**, BR-7715, DS-A0VM, E9-A1NA | 1,923 | 1,812 | 2,582 | 1,812 | 2,608 | 0.111 | 0.225 | 0.243 |
| F4 | 21-1077, 33-4532, A1-A0SH, **W5-AA30** | 1,156 | 999 | 1,801 | 999 | 1,725 | 0.143 | 0.268 | 0.254 |
| Pooled | | | | | | | 0.112 | 0.215 | 0.230 |

**Endpoint.** The primary endpoint is the clean-slide false-positive (FP) fraction: the fraction of the held-out clean slide's scored tissue flagged by the thresholding rule outside ground truth, pooled over the four folds by summing pixels. It is computed in two scored regions: the *entropy tissue region* (covered tissue by entropy-based detection, excluding teal ink identified by a fixed HSV rule: hue 150–210°, saturation ≥ 0.30, value ≤ 0.90 [12]) and the *saturation tissue region* (covered tissue by saturation-based detection, the convention of [12,13]). Sensitivity per artifact type is the fraction of annotated pixels flagged within the saturation tissue region.

### *2.5. Decision criteria*

For each comparison, a criterion on the pooled endpoint and on the number of folds agreeing was set before the comparison was evaluated, together with tolerances on sensitivity loss; each comparison was evaluated once. Criteria and outcomes are listed in Table 2.

**Table 2. Analyses, decision criteria and outcomes.** Rows follow the order of Section 3. Clean-slide FP: fraction of the held-out annotation-free slide's scored tissue flagged outside ground truth. Entropy region: covered tissue by entropy-based detection, excluding teal ink; saturation region: covered tissue by saturation-based detection. Original pool: saturation-Otsu tissue detection; curated pool: entropy-based tissue detection with marker ink removed after visual review.

| # | Question | Criterion | Data | Outcome | Criterion met |
|---|---|---|---|---|---|
| 1 | Does the curated pool reduce clean-slide FP? | Entropy-region clean FP of the curated pool (+ slide balance) ≤ 0.8 × original pool, and lower in ≥ 3 of 4 folds | k-fold, 4 folds, seed 42 | 0.1019 → 0.0158 (ratio 0.155); 4/4 folds | Yes |
| 2 | Same, in the saturation region | Saturation-region clean FP of the curated pool (+ slide balance) ≤ original pool (pooled) | as row 1 | 0.0768 → 0.0131; higher in F2 only | Yes |
| 3 | Which change carries the gain? | Ladder original pool → ink removed → entropy-based detection → slide balance; a step accounts for the gain if it carries ≥ 25% of the total reduction | k-fold, 4 folds, seed 42 | Tissue detection 101%; ink -1%; balance 0% | Yes (tissue detection step) |
| 4 | Does the gain come from the composition of the pool rather than its size? | Curated pool subsampled to the size of the original pool without marker ink keeps ≥ 75% of the gain | k-fold, 4 folds, seed 42 | 103% kept (0.0131; bar 0.0375) | Yes |
| 5 | Robust to the training seed? | Second seed: entropy-region ratio ≤ 0.8 and lower in ≥ 3 of 4 folds | k-fold, 4 folds, seed 7 | ratio 0.126; 4/4 folds | Yes |
| 6 | Is the effect graded by clear-space inclusion? | The GrandQC pool falls where its clear-space fraction places it relative to the other two pools, pooled and in ≥ 3 of 4 folds | k-fold, third tissue detector | as placed, pooled and in 4/4 folds (0.0087) | Yes |
| 7 | Is the gain paid for in sensitivity? | Saturation-region sensitivity not below the original pool by more than: air-bubble 0.05 (F2+F3), out-of-focus 0.03, penmark 0.05 | all k-fold arms | No arm outside tolerance; penmark and air-bubble rise | Yes |
| 8 | Does the effect hold for a feature-space detector? | UNI2-h nearest-neighbour detector: entropy-region ratio ≤ 0.8 and lower in ≥ 3 of 4 folds | k-fold, same folds | ratio 0.849; 3/4 folds | No |
| 9 | External: clean FP in our tissue region | SlideInspect saturation-region clean-control FP ratio ≤ 0.8 and 97.5% CI upper < 1 | SlideInspect, 55 clean controls | 0.780 [0.505, 1.002] | No (inconclusive) |
| 10 | External: clean FP in the annotator's tissue | Same criterion within the annotator's tissue mask | as row 9 | 0.806 [0.671, 0.988] | No (inconclusive) |
| 11 | External sensitivity cost | Out-of-focus sensitivity ≥ reference model - 0.05 | SlideInspect, 135 cases | -0.033 [-0.107, +0.020] | Yes |
| 12 | Is there a stain shift? | ≥ 50% of controls beyond the largest training-slide stain distance | SlideInspect vs TCGA pool | 11% | No |
| 13 | Does stain distance track errors? | Spearman rho(distance, clean FP) ≥ 0.3 | as row 12 | rho = 0.32 (organ-confounded) | Yes |

Notes. Inconclusive (rows 9, 10): the reduction was in the expected direction in both regions, but neither met the criterion for confirmation, and neither interval lay entirely above 1. Rows 12 and 13 characterize the external set rather than test the method. Row 6: only F3 separates the GrandQC and entropy-based pools clearly; in F2 and F4 the difference lies within seed variability (Fig. 3).

### *2.6. Feature-space detector*

To test whether the effect depends on the detector, we built a nearest-neighbour anomaly detector in the spirit of PatchCore [21] on UNI2-h [19] (same model revision and preprocessing as for the conditioning statistics). Each 1,024-pixel window is split into 4 × 4 crops of 256 pixels resized to 224; the last-layer patch tokens of each crop (16 × 16) are average-pooled to 4 × 4, giving a 16 × 16 grid of 1,536-dimensional vectors per window (one per 64 × 64 pixels). The memory bank of a fold contains all vectors of that fold's clean-pool patches (L2-normalized, exact search, no coreset), and the anomaly score is one minus the maximum cosine similarity to the bank. Maps use the same windows and grid as the diffusion detector and are scored by the same rule and endpoint.

### *2.7. External evaluation*

A model trained on the curated pool from all 16 slides (same recipe, checkpoint fixed at 1,000 steps) was compared once with the reference model on SlideInspect, with identical ingestion and noise draws. The reference model had been evaluated on SlideInspect once before, with the same ingestion code; the evaluation code of the present comparison reproduced that evaluation exactly (clean-control FP 0.226355 in the saturation tissue region). Two scored regions were used: the saturation tissue region, as in that earlier evaluation, and the annotator's tissue region (the manual tissue mask), on the 55 clean controls only. The endpoint is the clean-control false-positive fraction. Uncertainty was estimated by a stratified, paired case bootstrap (20,000 resamples, seed 0) with 97.5% intervals (Bonferroni correction for the two regions). A reduction in a region was regarded as confirmed if the ratio was ≤ 0.8 and the interval's upper bound < 1, as contradicted if the lower bound was > 1, and as inconclusive otherwise. Out-of-focus sensitivity (135 cases) was required not to fall by more than 0.05.

**Stain diagnostic.** For each SlideInspect image, the mean optical density of non-white pixels within the manual tissue mask was compared with the distribution of 300 random curated-pool patches by the Mahalanobis distance; the training slides' own distances define the within-TCGA range.

### *2.8. Descriptive analyses*

**Activation profile.** Patches from each fold's held-out slides that belong to the curated pool were divided into those rejected by saturation-based detection (clear-space) and those accepted by both methods (up to 32 per group and fold). Each batch was encoded and noised once (t = 800) and passed through the models trained on the two pools and their second-seed replicates; we recorded the output of every transformer block and the mean denoising error. The difference between the two pools is compared with the difference between seeds, which shares inputs and noise and is therefore a valid floor.

**Adapter spectra.** For each trained adapter, the singular values of the update of every module were computed from its low-rank factors.

### *2.9. Implementation*

Experiments ran on NVIDIA A100 GPUs (40 GB) in a Singularity container with PyTorch 2.5.1, diffusers 0.31.0, PEFT 0.19.1, timm 1.0.28, NumPy 2.1.2, SciPy 1.17.1 and LazySlide 0.12.0. Code, pools, fold definitions and evaluation outputs are available at Zenodo (doi:10.5281/zenodo.23016733).

## 3. Results

### *3.1. What saturation-based tissue detection leaves out of the clean pool*

On the annotation-free grid of the 16 training slides, per-slide saturation–Otsu tissue detection accepted 2,185 patches and entropy-based detection 3,071. The two methods agreed on 2,078 patches; entropy-based detection added 993 patches that saturation-based detection rejected and dropped 107 that it accepted. The differences arise from two distinct failure modes (Fig. 1). On the three slides carrying thick marker ink (TCGA-85-8052, TCGA-86-8076, TCGA-BR-7715), the per-slide Otsu threshold was high (97–111 of 255, against 39 and 84 on two reference slides), and the saturation-based mask kept the ink and the most intensely stained tissue while excluding ordinary tissue (Fig. 1a; Supplementary Fig. S1). Removing teal ink pixels before thresholding did not lower the threshold (it rose by 2–6 levels), and red ink cannot be separated from eosin by color, so the share of the ink in setting the threshold could not be isolated. All 107 patches accepted only by saturation-based detection lie on these slides, together with 227 of the patches it rejected, which include ordinary, dense tissue. On the other 13 slides, saturation-based detection never accepted a patch that entropy-based detection rejected; it only excluded tissue, 766 patches, mostly on lung (TCGA-55-A492 alone contributed 451; Fig. 1b) and breast slides. In a random sample of 64 of these patches, stratified by slide and inspected visually, about half showed sparse tissue with large clear spaces (alveolar parenchyma, loose stroma, adipose tissue; Fig. 1c, E–G), about a third lay on the tissue boundary with close to half the box on glass (a difference in where the two masks draw the edge), and the remainder were tissue filling the box but weakly stained, including the frozen section. On the inked slides, 30 of a random sample of 32 excluded patches showed ordinary tissue without ink, including dense carcinoma whose saturation-based tissue fraction was below 0.05 (Fig. 1c). As a result, 155 of the 2,185 patches of the original pool (7.1%) carried marker ink on visual review, while stretches of normal tissue on the same slides were excluded. The original pool was also dominated by a single densely stained slide (TCGA-33-4532; 776 of 2,185 patches, 35.5%). The curated pool (entropy-based detection, marker ink removed after visual review) contains 2,969 patches.

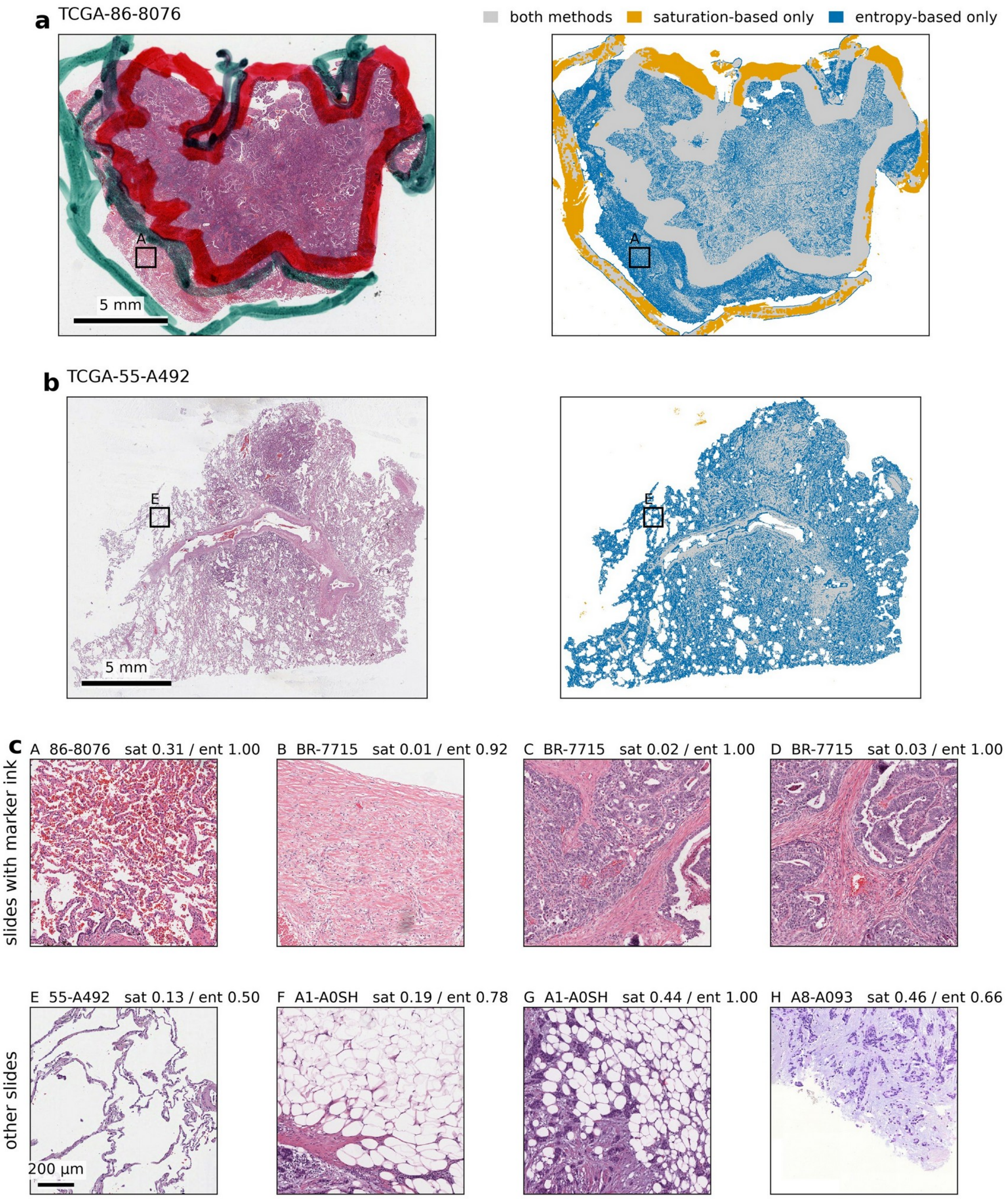

**Fig. 1.** What saturation-based tissue detection leaves out of the clean pool. (a) TCGA-86-8076, a lung slide with thick marker ink: thumbnail (left) and agreement between saturation-based tissue detection (per-slide Otsu threshold on the HSV saturation of the thumbnail) and entropy-based tissue detection (right). The per-slide Otsu level is 97 of 255; the saturation-based mask follows the red and teal ink and the most intensely stained tissue, and misses 34% of the tissue found by entropy-based detection. (b) TCGA-55-A492, lung with large clear spaces (Otsu level 39): saturation-based detection misses 52% of the tissue found by entropy-based detection and adds almost none of its own. (c) Patches (1,024 × 1,024 pixels at 10×, about 1 mm across; scale bar in E) rejected by saturation-based and admitted by entropy-

based detection, all in the curated pool; values give the fraction of the patch classified as tissue by each method (admission threshold 0.5). Top, slides with marker ink: ordinary tissue without ink, including carcinoma (C, D). Bottom, other slides: alveolar parenchyma (E), adipose tissue (F, G) and weakly stained tissue at the tissue boundary (H). Boxes in (a) and (b) mark patches A and E.

### *3.2. Changing the tissue detector removes most clean-slide false positives*

On the four held-out clean slides, the curated pool, trained with per-slide balancing, reduced the fraction of tissue flagged outside ground truth from 0.1019 to 0.0158 in the entropy tissue region (ratio 0.155; lower in all four folds) and from 0.0768 to 0.0131 in the saturation tissue region (Table 2, rows 1 and 2). The attribution ladder located the entire gain at the change of tissue detector (Table 2, row 3; Fig. 2): removing marker ink from the original pool left clean-slide FP unchanged (0.1030), replacing saturation-based with entropy-based detection brought it to 0.0157, and per-slide balancing added nothing (0.0158; its per-fold effects went in both directions and were of the same order as the difference between training seeds). All later comparisons therefore use the curated pool without balancing. In every fold the drop occurred at the tissue-detection step; it was largest on the pale, fat-rich breast slide TCGA-A8-A093, from 0.470 to 0.068 at that step.

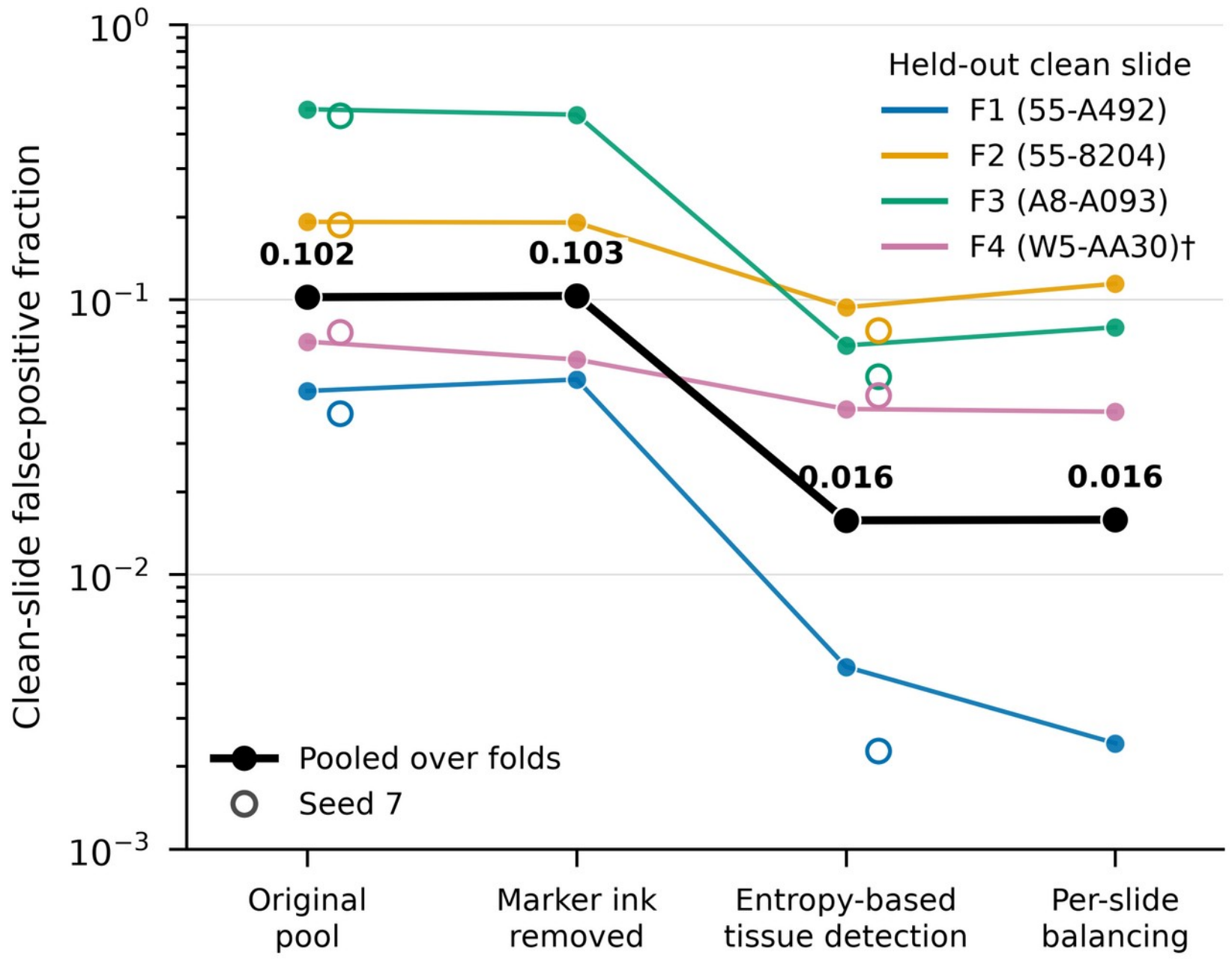

**Fig. 2.** Where the gain comes from. Clean-slide false-positive fraction (entropy tissue region, log scale) on the held-out clean slide of each fold as the clean pool is changed one step at a time: original pool (saturation-based tissue detection), marker ink removed, entropy-based tissue detection (curated pool), and per-slide balancing. Colored lines, individual folds; black, pooled over folds (values shown); open circles, second training seed for the original and curated pools. Removing marker ink and balancing slides leave the pooled value unchanged; the change of tissue detection lowers it from 0.103 to 0.016 and lowers it in every fold. †Frozen section.

Two controls support this reading. The curated pool subsampled to the size of the original pool without marker ink kept 103% of the gain (0.0131; Table 2, row 4), so the effect comes from what enters the pool, not from how much. Retraining on both pools with a second seed reproduced the effect (0.0927 to 0.0117, ratio 0.126, lower in all folds; Table 2, row 5); the seed-to-seed spread of the pooled endpoint (0.009 for the original pool, 0.004 for the curated pool) is about a tenth of the effect.

In the saturation tissue region the gain was concentrated on TCGA-A8-A093 (0.453 to 0.060) and TCGA-W5-AA30 (0.034 to 0.023); TCGA-55-A492 was already near zero (0.0007 to 0.0000), and on TCGA-55-8204 the false-positive fraction rose slightly (0.053 to 0.060; 0.050 without per-slide balancing). This is consistent with that region itself excluding much of the tissue the curated pool adds.

### *3.3. The effect is graded by the clear-space tissue a detector includes*

If the mechanism is the inclusion of clear-space tissue, a third tissue detector should fall where its clear-space fraction places it. The clear-space fraction (mean fraction of near-white pixels per patch) needs no labels and no training; we computed it for each fold's pool under saturation-based, entropy-based and GrandQC tissue detection [4] and compared it with the clean-slide FP of the model trained on each pool (Table 2, row 6). Pooled over folds, the clear-space fractions were 0.112, 0.215 and 0.230 and the clean-slide FP 0.1030, 0.0157 and 0.0087, respectively (Fig. 3). The GrandQC pool fell where its clear-space fraction placed it, in the pooled result and in all four folds.

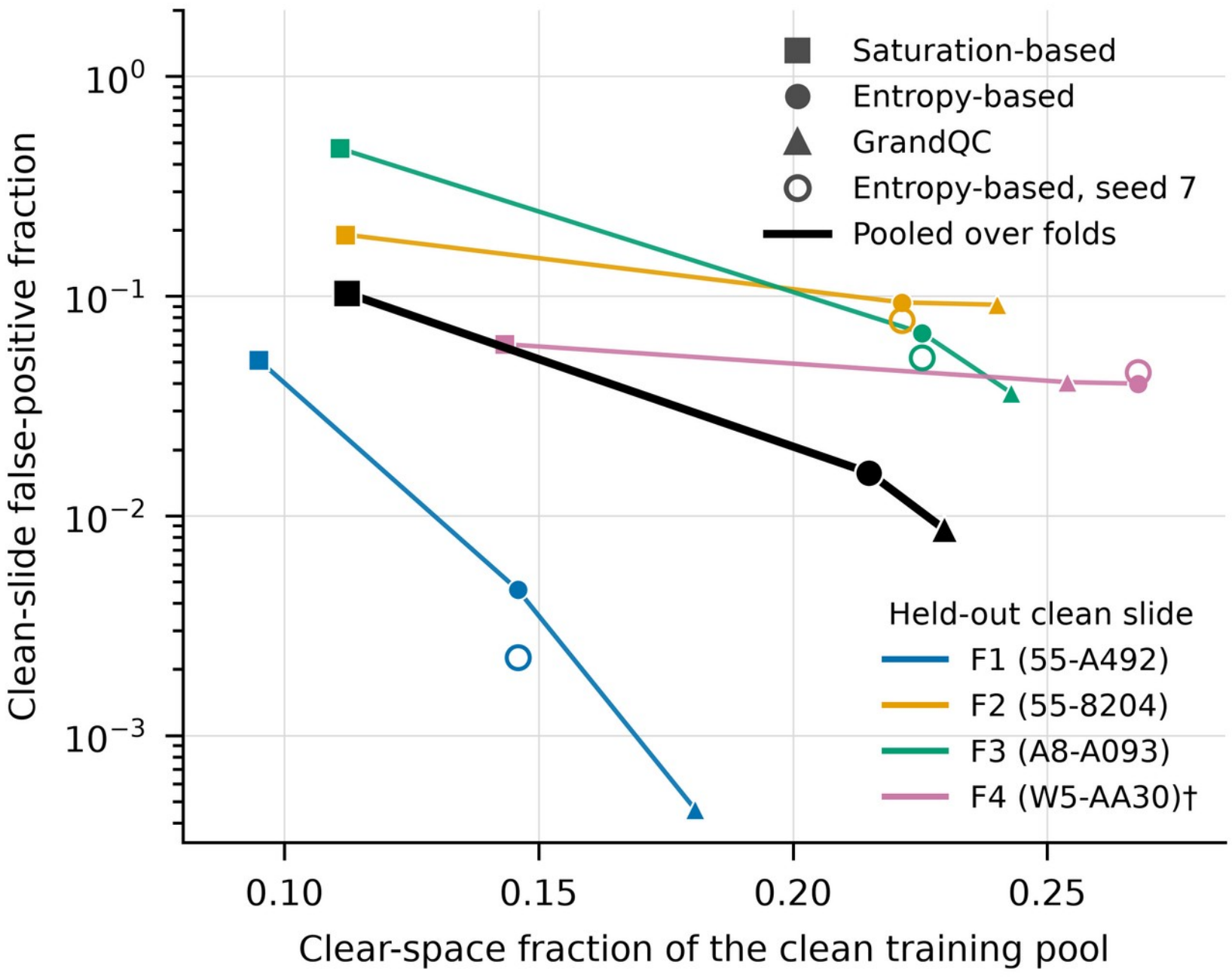

**Fig. 3.** The gain is graded by the clear-space tissue the pool contains. Clean-slide false-positive fraction (entropy tissue region, log scale) against the clear-space fraction of the clean training pool (mean fraction of near-white pixels per patch; computed without labels or training), for pools built with three tissue detection methods: saturation-based (marker ink removed), entropy-based (curated pool) and GrandQC. Colored lines, individual folds; black, pooled over folds; open circles, curated pool with a second training seed. Pooled, the clear-space fractions were 0.112, 0.215 and 0.230 and the false-positive fractions 0.103, 0.016 and 0.009. Nearly all of the reduction occurs between saturation-based detection and the other two methods; the difference between entropy-based and GrandQC pools is clear on TCGA-A8-A093 (F3), small on TCGA-55-A492 (F1), where both are below 0.005, and within seed variability on TCGA-55-8204 (F2) and TCGA-W5-AA30 (F4). †Frozen section.

The relation is monotone and strongly non-linear: almost all of the reduction occurs between saturation-based detection and the other two. The difference between the entropy-based and GrandQC pools is small; it is clear on TCGA-A8-A093 (0.068 to 0.036), near zero on TCGA-55-A492, where both are already below 0.005, and within seed variability on TCGA-55-8204 and TCGA-W5-AA30 (Fig. 3, open circles). The pooled difference (−0.0070) is about 1.7 times the seed spread of the entropy-based pool.

### *3.4. The gain is not paid for in sensitivity*

No model fell outside the sensitivity tolerances we set (Table 2, row 7; Table 3). In the saturation tissue region, pooled over folds, sensitivity moved from the original to the curated pool as follows: out-of-focus 0.656 to 0.670, fold 0.137 to 0.149, penmark 0.521 to 0.584 and air bubble 0.086 to 0.102; the GrandQC pool gave 0.660, 0.140, 0.591 and 0.115. Penmark and air-bubble sensitivity rose with every change of tissue detector.

**Table 3. Clean-slide false positives and sensitivity of every diffusion-detector model, pooled over the four held-out folds.** Clean-slide FP in the entropy and saturation tissue regions; sensitivity per artifact type (fraction of annotated pixels flagged) in the saturation tissue region. Annotated pixels: out-of-focus 1,724,384, fold 221,697, penmark 3,019,023, air bubble 6,188,216; air bubbles occur only in folds F2 and F3.

| Pool | Seed | Clean FP, entropy region | Clean FP, saturation region | Out-of-focus | Fold | Penmark | Air bubble |
|---|---|---|---|---|---|---|---|
| Original pool | 42 | 0.1019 | 0.0768 | 0.656 | 0.137 | 0.521 | 0.086 |
| Original pool, marker ink removed | 42 | 0.1030 | 0.0729 | 0.690 | 0.128 | 0.556 | 0.094 |
| Curated pool | 42 | 0.0157 | 0.0116 | 0.670 | 0.149 | 0.584 | 0.102 |
| Curated pool, per-slide balancing | 42 | 0.0158 | 0.0131 | 0.645 | 0.149 | 0.630 | 0.116 |
| Curated pool, subsampled to original size without ink | 42 | 0.0131 | 0.0090 | 0.661 | 0.131 | 0.594 | 0.133 |
| GrandQC pool | 42 | 0.0087 | 0.0084 | 0.660 | 0.140 | 0.591 | 0.115 |
| Original pool | 7 | 0.0927 | 0.0729 | 0.672 | 0.128 | 0.509 | 0.083 |
| Curated pool | 7 | 0.0117 | 0.0100 | 0.674 | 0.139 | 0.590 | 0.108 |

### *3.5. The curated pool teaches the model to reconstruct clear-space tissue*

To connect the endpoint to the model's behavior, patches from the held-out slides were passed, with the same latent and noise, through the models trained on the two pools and through their second-seed replicates (descriptive analysis). On clear-space patches that saturation-based detection rejects, the curated pool lowered the mean denoising error at $t = 800$ by 0.039, about seven times the seed-to-seed difference (0.005–0.006); on patches accepted by both methods the reduction was 0.007 (seed difference 0.002–0.003). Internal activations differed between the two pools beyond the seed floor at every transformer block except the first, more on clear-space than on dense patches (1.2–1.65 times), with the largest pool-specific difference in the last blocks.

The low-rank adapters themselves are nearly rank-2 (two singular values carry 90% of each module's update), and distances between adapters in weight space do not track function:

removing marker ink moved the adapters almost as much as changing the tissue detector, while only the latter changed the endpoint (Supplementary Note S1).

### *3.6. The effect depends on the detector*

A nearest-neighbour anomaly detector on UNI2-h patch features [19], using the same pools, folds, windows and scoring, did not reproduce the effect at the size required by our criterion: its pooled clean-slide FP fell from 0.1062 to 0.0902 (ratio 0.849, lower in three of four folds; Table 2, row 8). The two detectors failed on different slides. The diffusion detector's largest failure was the fat-rich TCGA-A8-A093, which the curated pool repaired; the nearest-neighbour detector handled that slide with either pool (0.143 and 0.137) but flagged 98% of TCGA-W5-AA30, the only frozen section in the set, where the diffusion detector flagged 0.07. Its FP on TCGA-55-8204 (0.57 and 0.48) is unexplained. Sensitivities are not compared across detectors, whose flagged fractions at the shared thresholding rule differ widely.

### *3.7. External test: the right direction, not confirmed, and not explained by stain*

On SlideInspect [17], a single comparison of the model trained on the curated pool (all 16 slides) against the reference model lowered the clean-control FP from 0.2264 to 0.1766 in the saturation tissue region (ratio 0.780, 97.5% confidence interval [CI] 0.505–1.002) and from 0.2305 to 0.1857 in the annotator's tissue region (ratio 0.806, 0.671–0.988); neither met our criterion for confirmation (Table 2, rows 9 and 10; Table 4). Out-of-focus sensitivity stayed within the tolerance (0.818 to 0.785; difference −0.033, CI −0.107 to +0.020; Table 2, row 11). Descriptively, pooled F1 rose from 0.378 to 0.403 and fold sensitivity from 0.431 to 0.546; among organs with more than one control, the largest clean-control reductions were in prostate (0.585 to 0.280) and, within the annotator's region, breast (0.317 to 0.206), with none in liver (0.017 to 0.025; Table 4b).

**Table 4. External evaluation on SlideInspect.** H&E images (n = 289; 55 clean controls without artifact annotation), reference model versus the model trained on the curated pool, same thresholding rule and noise draws. Ratios and differences with 97.5% paired bootstrap intervals (20,000 resamples).

*(a) Overall*

| Measure | Reference model | Curated pool | Ratio or difference [97.5% CI] |
|---|---|---|---|
| Clean-control FP, saturation region | 0.226 | 0.177 | 0.780 [0.505, 1.002] |
| Clean-control FP, annotator's tissue region | 0.230 | 0.186 | 0.806 [0.671, 0.988] |
| Out-of-focus sensitivity (135 images) | 0.818 | 0.785 | −0.033 [−0.107, +0.020] |

| Measure | Reference model | Curated pool | Ratio or difference [97.5% CI] |
|---|---|---|---|
| Mounting-artifact sensitivity (32 images) | 0.935 | 0.914 | |
| Fold sensitivity (174 images) | 0.431 | 0.546 | |
| Pixel F1, saturation region | 0.378 | 0.403 | |

*(b) Clean-control FP by organ*

| Organ | Controls | Saturation region, reference | Saturation region, curated | Annotator's region, reference | Annotator's region, curated |
|---|---|---|---|---|---|
| Breast | 21 | 0.367 | 0.312 | 0.317 | 0.206 |
| Liver | 10 | 0.017 | 0.025 | 0.025 | 0.033 |
| Colon | 9 | 0.280 | 0.221 | 0.306 | 0.322 |
| Lung | 7 | 0.110 | 0.073 | 0.158 | 0.131 |
| Prostate | 7 | 0.585 | 0.280 | 0.563 | 0.330 |
| Kidney | 1 | 0.275 | 0.005 | 0.323 | 0.040 |

The curated pool therefore recovers only a small part of the cross-center loss: a reduction of about 20%, against more than sixfold within TCGA. Stain does not explain the remainder: the median stain distance of the 55 controls to the training pool (2.28) lay within the range of the training slides themselves (0.53–5.00), and only 11% of controls exceeded the largest training-slide distance (Table 2, row 12). Stain distance correlated moderately with clean FP ($\rho = 0.32$; Table 2, row 13), but the correlation is confounded by organ: colon and prostate had both the largest distances and, in the annotator's tissue region, the highest residual FP (Table 4b); in the saturation region, breast had the highest.

## 4. Discussion

### *4.1. The clean pool defines normal*

A one-class detector can only call normal what its training pool contained. Our results show how directly this plays out through a preprocessing step usually considered neutral. Per-slide saturation–Otsu tissue detection, close to the default of the widely used CLAM preprocessing pipeline (saturation channel of the HSV thumbnail, median filter, fixed or Otsu threshold) [22], excludes normal tissue with large clear spaces: adipose tissue, alveolar parenchyma and loose stroma. Excluded from the pool, this tissue lies outside the learned distribution, and the detector flags it on unseen slides. Replacing the tissue detector removed most of these false positives, the gain came from what entered the pool rather than how much, and a third detector fell where the clear-space fraction of its pool placed it. The model trained on the curated pool reconstructs clear-space tissue markedly better than the model trained on the original pool, while its reconstruction of dense tissue barely changes (Section 3.5), which ties the endpoint to the mechanism.

The blind spot is architectural rather than chromatic. Saturation-based detection is often described as failing on "pale" tissue; in our slides the excluded tissue was not pale so much as sparse: in a visually reviewed sample, about half was lumen with thin walls or septa, about a third lay on tissue boundaries, and weakly stained tissue filling the box was a minority (Section 3.1). This matters for remedies: stain normalization or color augmentation would not bring it into the pool, whereas a tissue detector that responds to texture does. It agrees with a comparison of channel reductions for tissue segmentation, in which local entropy separated tissue from background best and grayscale worst [23].

The relation between clear-space fraction and false positives is strongly non-linear (Fig. 3). Nearly all of the gain is obtained once the pool contains clear-space tissue at all; adding more, as the GrandQC masks do, helps mainly on the one slide rich in it. In practice this means that the choice among reasonable tissue detectors matters much less than avoiding one that systematically excludes clear-space tissue.

The same mechanism has consequences beyond detection. Artifact restoration methods that rewrite the regions a detector flags [5,24–26] would, driven by a detector with this blind spot, rewrite normal fat and lung parenchyma. Such methods also learn normal tissue from regions a quality-control tool declares artifact-free (GrandQC masks in [5]), so the question studied here, which tissue that tool admits, applies to them as well; we did not test it. And because the reference pipeline also used saturation-based detection to define the scored region [14], the benchmark and the training pool shared the blind spot, so that published performance figures were measured largely outside it.

### *4.2. Why the effect depends on the detector*

The nearest-neighbour detector on UNI2-h features did not show the effect at the size required by our criterion, and it failed on different slides than the diffusion detector (Section 3.6). One reading, which we did not test, is that the two detectors define normal at different levels. The diffusion detector judges whether a patch can be reconstructed from noise, a low-level property that transfers across tissue types learned during pretraining but suffers on textures absent from the pool. The nearest-neighbour detector judges proximity in a semantic feature space, in which clear-space tissue may already lie close to patches in the pool, while a different preparation (the frozen section in our set) moves whole slides away from it.

Whatever the explanation, the practical conclusion is that a "good" clean pool is relative to the detector: a pool curated for one family of detectors should be audited again before it is reused for another.

### *4.3. The cross-center gap*

On SlideInspect, the curated pool moved clean-control false positives in the expected direction by about 20%, in both scored regions, without meeting our criterion for confirmation. Within TCGA the same change reduced them more than sixfold. Most of the cross-center loss therefore has another cause, and stain is not it: the external images fell within the stain range of the training slides. In the annotator's tissue region the residual false positives were highest in colon and prostate, organs absent from a pool that is three-quarters lung (76% of patches); but liver, also absent, had few, and breast, which is in the pool, had the highest residual in the saturation region, so organ coverage alone does not explain the pattern; the storage of SlideInspect at 5× is a further candidate. These observations are post hoc and are offered as hypotheses. Testing the organ-coverage hypothesis requires a pool that includes the missing organs and an external set that the hypothesis was not derived from, which is a natural next step.

### *4.4. Practical recommendations*

For groups building clean pools for one-class quality control, our results suggest five inexpensive practices. (i) Do not build the pool with per-slide saturation–Otsu tissue detection; entropy-based detection or the public GrandQC masks for TCGA [4] gave pools with a sixth to a twelfth of the clean-slide false positives. (ii) Report the clear-space fraction of the pool, a label-free statistic that took minutes to compute, needs no training, and ordered three pools as their clean-slide false positives. (iii) On slides with thick marker ink, review the pool for ink: a per-slide threshold can keep the ink while excluding ordinary tissue. (iv) Report separately the tissue detection used to build the pool and the one used to define the scored region; they need not coincide, and when they share a blind spot the evaluation cannot reveal it. (v) Evaluate on held-out slides without artifacts and report their false-positive fraction, which is the endpoint most directly affected by the pool.

### *4.5. Comparing fine-tuned adapters*

One methodological observation may help studies that compare fine-tuned adapters. Distances between low-rank adapters in weight space did not track their function in our models, largely because adapters trained with different seeds start from different random projections (Supplementary Note S1). Functional comparisons on shared inputs and noise, as in Section 3.5, avoid this.

### *4.6. Limitations*

The development evidence rests on 16 slides, and the primary endpoint on the four clean slides held out one per fold; the effect is consistent across folds and seeds, but the number of slides limits how far it generalizes. The slides come from TCGA, which is part of the backbone's pretraining corpus, so all cross-validation results are development-tier; the only external set is available at 5× and yielded an unresolved result. We trained two seeds per pool, and the difference between entropy-based and GrandQC tissue detection is within seed variability on two of four folds. One held-out clean slide is a frozen section. Marker ink was reviewed by eye by one reader under an ink-only rule. The detector-dependence finding rests on two detector families, and sensitivities are not compared across them. Finally, the mechanism analyses of Section 3.5 are descriptive.

## 5. Conclusion

For a one-class artifact detector, tissue detection is not a neutral preprocessing step: it decides what the model learns as normal. Per-slide saturation–Otsu detection excludes normal tissue with large clear spaces, and a diffusion-based detector trained on such a pool flags that tissue as artifact. Replacing the tissue detector removed most false positives on held-out clean slides without loss of sensitivity, in a graded way that followed the clear-space fraction of the pool. The effect is specific to the detector family and, across centers, points in the right direction without being confirmed. Choosing and reporting how the clean pool is built should be part of the standard description of one-class quality control methods.

## CRediT authorship contribution statement

**Konstantinos Moutselos:** Conceptualization, Methodology, Software, Formal analysis, Investigation, Data curation, Visualization, Writing – original draft. **Ilias Maglogiannis:** Conceptualization, Supervision, Funding acquisition, Writing – review & editing.

## Acknowledgments

Computing resources were provided by the University of Piraeus, NVIDIA DGX GPU Cluster. We thank Massimo Salvi for information on the SlideInspect release.

## Ethics Statement and Patient Consent

This study used only publicly available, de-identified images: whole-slide images from The Cancer Genome Atlas with artifact annotations from the AIRAQC benchmark, and the public release of the SlideInspect dataset. No new tissue, images or patient data were collected, and the authors had no access to identifying information. Patient consent and approvals for data collection were obtained by the original data providers; separate ethics approval and patient consent were therefore not required for this secondary analysis.

## Data availability

The whole-slide images are available from the TCGA data portal, the artifact annotations from the AIRAQC authors [15], the GrandQC tissue masks for TCGA from their public deposit (doi:10.5281/zenodo.14041578, CC BY-NC-SA 4.0) [4], and the SlideInspect images from Kaggle (artemis90/slideinspect) under its academic, non-commercial terms. The clean pools, fold definitions, marker-ink labels, evaluation outputs, the low-rank adapters of all models, and the code for pool construction, training, evaluation, figures and tables are available at Zenodo, doi:10.5281/zenodo.23016733 [27].

## Declaration of generative AI and AI-assisted technologies in the manuscript preparation process

During the preparation of this work the authors used Claude (Anthropic) in order to draft and edit the manuscript text and write the analysis, figure and table scripts released with the paper. All experiments were run by the authors, and all reported numbers were checked against the evaluation outputs. After using this tool, the authors reviewed and edited the content as needed and take full responsibility for the content of the published article.

# Supplementary material

*The Effect of Tissue Detection on False Positives of Diffusion-Based Artifact Detection in Histopathology*

## Supplementary Note S1. Spectra and weight-space distances of the low-rank adapters

**Spectra.** For every trained model we computed, for each of the 224 adapted modules, the singular values $\sigma_i$ of the low-rank update

$$\Delta W = \frac{\alpha}{r} BA$$

obtained from the r × r core after QR decomposition of the two factors (rank r = 16, scaling α = 16). Across all 24 models (six per fold × four folds), the median number of singular values needed to reach 90% of a module's squared Frobenius norm was 2, and the mean participation ratio

$$\mathrm{PR} = \frac{\left(\sum_i \sigma_i^2\right)^2}{\sum_i \sigma_i^4}$$

was 2.01–2.15. The adapters are therefore effectively of rank 2, although they were trained with rank 16.

**Distances between adapters do not track function.** We compared two adapters a and b by the relative difference of their updates. For module m it is

$$d_m = \frac{||\Delta W_{a,m} - \Delta W_{b,m}||_F}{\sqrt{||\Delta W_{a,m}||_F \; ||\Delta W_{b,m}||_F}}$$

and over all modules the average weighted by the geometric mean of the two norms

$$D = \frac{\sum_m ||\Delta W_{a,m} - \Delta W_{b,m}||_F}{\sum_m \sqrt{||\Delta W_{a,m}||_F \; ||\Delta W_{b,m}||_F}}$$

Between models trained with the same seed on different pools, relative differences were 0.21–0.39. Between models trained on the same pool with different seeds, they were 1.08–1.10 in every fold, and 0.99–1.18 at every transformer block of every fold; the two updates are nearly orthogonal. Yet the same pairs behave almost identically: clean-slide false-positive fractions of 0.0157 and 0.0117 for the curated pool (Table 3), and seed differences in denoising error of 0.002–0.006 (Section 3.5). The reason is the initialization: in LoRA the matrix A is drawn at random from the training seed, so adapters trained with different seeds start from different random projections; the near-orthogonal updates are consistent with functionally similar solutions expressed in different coordinates. A seed-to-seed distance in weight space is therefore not a valid noise floor for comparing adapters.

Among same-seed pairs, where this problem does not arise, distances still did not follow the endpoint. Removing marker ink moved the adapters by 0.26 on average (range 0.23–0.32 over folds) and left clean-slide false positives unchanged (0.102 to 0.103); changing the tissue detector moved them by 0.32 (0.23–0.36) and lowered false positives from 0.103 to 0.016; replacing entropy-based by GrandQC detection moved them by 0.22 (0.21–0.23) and lowered false positives further, to 0.009. Functional comparisons on shared inputs and noise (Section 3.5) avoid both problems.

**Update norm.** One weight-space quantity did differ consistently with the pool. The total Frobenius norm of the update was larger for pools built with entropy-based or GrandQC tissue detection than for pools built with saturation-based detection, in every fold and with both seeds (curated vs original pool: 2–7% larger with seed 42, 1–6% with seed 7). The difference is modest and, in two folds, of the same size as the seed-to-seed difference of the curated pool; we report it without interpretation.

## Supplementary Fig. S1

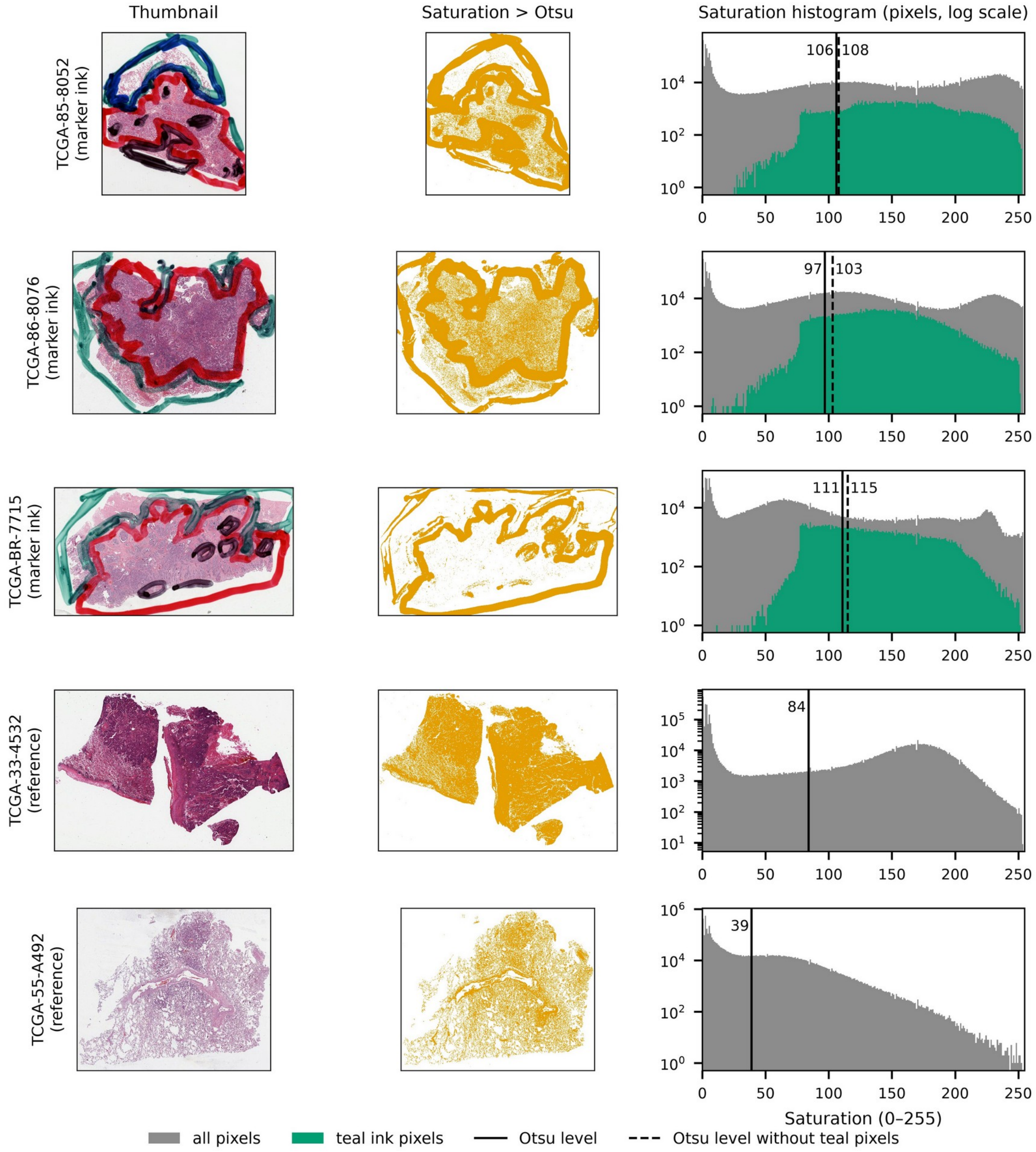

**Supplementary Fig. S1.** Per-slide saturation–Otsu thresholds on the three training slides with thick marker ink (top three rows) and two reference slides without ink. Left, slide thumbnail; middle, saturation-based tissue mask (HSV saturation above the per-slide Otsu level); right, histogram of thumbnail saturation (log scale) for all pixels (gray) and for pixels of the teal ink mask (green; hue 150–210°, saturation ≥ 0.30, value ≤ 0.90), with the Otsu level (solid) and the Otsu level recomputed without teal pixels (dashed). On the inked slides tissue fills most of the thumbnail and the threshold (97–111 of 255) falls within the tissue, so the mask keeps the ink and the most intensely stained tissue and excludes paler tissue; excluding teal pixels does not lower it. TCGA-85-8052 also carries blue and black ink; red ink cannot be separated from eosin by color. On the reference slides the threshold (84 and 39) lies between the glass and tissue modes of the histogram.